\documentclass[aps,amssymb,amsmath,pra,reprint,noshowpacs]{revtex4-1}
\usepackage{times}
\usepackage{amssymb,amsmath,graphicx}
\usepackage[usenames]{color}
\usepackage{bbold}
\usepackage{siunitx}
\usepackage{braket}
\usepackage{bbm}

\usepackage{hyperref}
\hypersetup{colorlinks=true, citecolor=blue, urlcolor=blue, linkcolor=blue}

\usepackage[T1]{fontenc}
\usepackage[utf8]{inputenc}
\usepackage[english]{babel}

\usepackage[normalem]{ulem}
\usepackage{lineno}

\newcommand{\NA}{\mathrm{N \mskip-0.5\thinmuskip A}}
\newcommand{\vect}[1]{\mathbf{#1}}

\newcommand{\imu}{\text{\rm i}}
\newcommand{\diff}{\text{d}}
\usepackage{textcomp}

\newcommand{\commentOut}[1]{}
\usepackage{scalefnt}   

\newcommand{\affil}{Photonics Laboratory, ETH Zürich, CH-8093 Zürich, Switzerland}

\begin{document}
\scalefont{1.05}
\title{Optimal Position Detection of a Dipolar Scatterer in a Focused Field}
\author{Felix Tebbenjohanns}
\affiliation{\affil}
\author{Martin Frimmer}
\affiliation{\affil}
\homepage{http://www.photonics.ethz.ch}
\author{Lukas Novotny}
\affiliation{\affil}


\begin{abstract}
We theoretically analyze the problem of detecting the position of a classical dipolar scatterer in a strongly focused optical field. We suggest an optimal measurement scheme and show that it resolves the scatterer's position in three dimensions at the Heisenberg limit of the imprecision-backaction product. We apply our formalism to levitated-optomechanics experiments and show that backscattering detection provides sufficient information to feedback-cool the particle's motion along the optical axis to a phonon occupancy below unity under realistic experimental conditions.
\end{abstract}
\date\today

\maketitle

\section{Introduction}
Soon after the advent of the laser, scientists have started to investigate and harness the forces arising from the interaction of light and matter~\cite{Phillips1998}. Of outstanding practical significance are the works on optically trapping micron and sub-micron sized particles in dilute gas and in liquid, which led to the optical tweezer, an indispensable tool in the life sciences~\cite{Ashkin1980,Ashkin1986,Ashkin2006}. Interestingly, optical trapping in vacuum has seen a revival in recent years in the wake of optomechanics rising as an intensely researched subfield of physics~\cite{Chang2010,Romero-Isart2011a,Li2011,Gieseler2012}. 
Optomechanics strives to control mechanical motion using the forces of light~\cite{Aspelmeyer2014}. The experimental platforms harnessed by optomechanics are high-quality mechanically tethered oscillators, whose position is read out optically. The limits of this measurement process are understood since the early theoretical works in the context of designing interferometer-based gravitational-wave detectors~\cite{Caves1980,Caves1987,Braginsky1992}. 
In the canonical optomechanical setup, the position of a mirror reflecting a beam of light is encoded into the field's phase and is read out interferometrically~\cite{Bowen2015}. At the same time, the radiation-pressure fluctuations of the probe light influence the mechanical motion of the mirror. Accordingly, the measurement of the mirror's position inevitably entails a perturbation of its momentum~\cite{Purdy2013,Teufel2016,Peterson2016}. In the limit of perfect detection efficiency, and in the absence of dephasing mechanisms besides radiation pressure shot noise, the product of the measurement imprecision and the measurement backaction satisfies the Heisenberg relation with equality~\cite{Clerk2010}. In recent years, harnessing enhancement effects provided by optical cavities, the optomechanics community has pushed the position detection of mechanically clamped oscillators to operate essentially at the Heisenberg limit~\cite{Rossi2018}. 

Levitated optomechanical systems are experimental test-beds complementary to mechanically clamped systems~\cite{Romero-Isart2011a,Yin2013a}. One particularly intriguing aspect of optical levitation is the fact that the potential experienced by the mechanical oscillator can be tuned by shaping the optical field distribution. Thus, optically levitated systems may enable tests of our understanding of physics at a very fundamental level and in parameter regimes currently inaccessible~\cite{Geraci2010,Arvanitaki2013,Moore2014,Rider2016}. The current forerunner in the quest for bringing an optically levitated system to the quantum regime is a sub-wavelength sized dielectric nanoparticle trapped in the focus of a single laser beam~\cite{Gieseler2012,Jain2016,Rodenburg2016,Vovrosh2017}. The center-of-mass motion of such a particle has been cooled to a population of a dozen phonons using active feedback cooling in a system operating three orders of magnitude from the Heisenberg limit of maximum detection efficiency~\cite{Tebbenjohanns2019}. At the current stage, pushing optically levitated systems into the quantum regime relies on improving the detection efficiency for the position of a dipolar scatterer in a focused light field~\cite{Gittes1998,Dawson2019}.  
To this end, remarkable progress has been made in recent years to couple optically levitated nanoparticles to optical resonators~\cite{Kiesel2013, Millen2015, Fonseca2016, Windey2019, Delic2019}, in efforts inspired by the atomic physics community~\cite{Qamar2000,Vuletic2001,Murch2008,Koch2010}. 
Interestingly, while sophisticated detection systems using optical cavities are under development, the question of the reachable position-detection efficiency for a dipolar scatterer in a single-beam optical trap has remained unanswered. 

In this paper, we theoretically analyze the problem of how to optimally measure the position of an isotropic dipolar scatterer in a focused light field. We derive a scheme to detect the motion of a nanoparticle optically trapped in a focused light field which operates strictly at the Heisenberg limit of optimal detection efficiency. Furthermore, we analyze the efficiencies of detection schemes currently employed by the optical-levitation community. Our results show that a simple backscattering configuration provides a detection efficiency for the oscillation mode along the optical axis of more than 60\% of the Heisenberg limit. Accordingly, active feedback cooling of a levitated nanoparticle's center-of-mass motion to a phonon occupancy below unity is feasible in a single-beam optical trap.

\begin{figure}[b]
\includegraphics[width=\columnwidth]{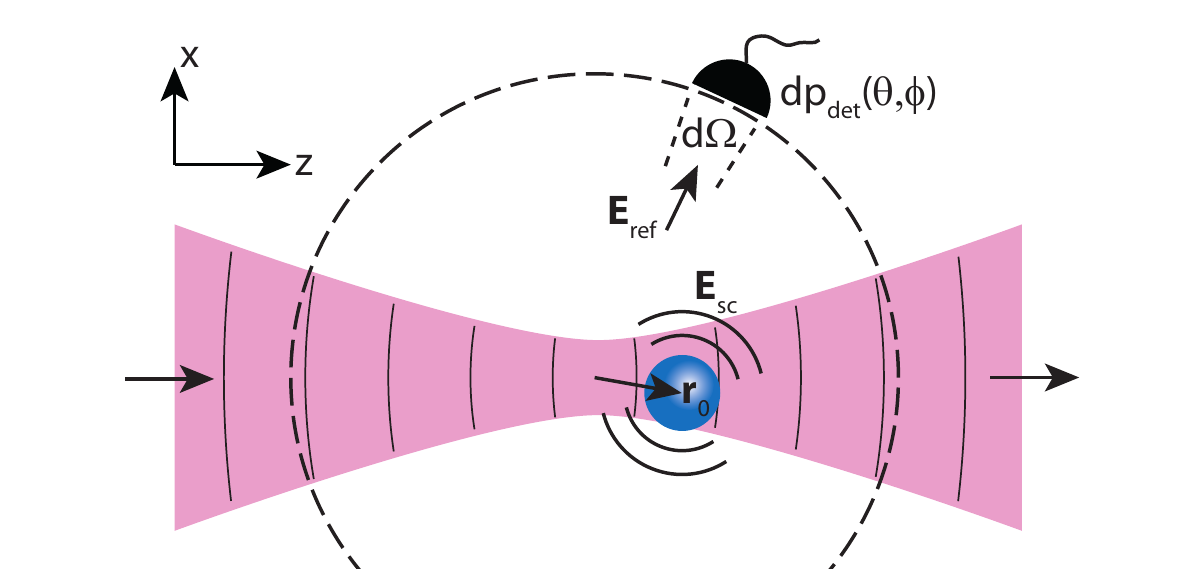}
\caption{Sketch of ideal measurement scheme. An isotropic dipolar scatterer is located at position $\vect{r}_0$ relative to the origin, which coincides with the focal point of a beam of light travelling from left to right. An array of detectors (only one depicted for clarity, each covering a solid angle $\diff\Omega$) is covering a sphere centered on the origin with radius much larger than the wavelength. A reference field $\mathbf{E}_{\rm ref}$ is added at the detector for homodyne detection of the scattered field $\vect{E}_\text{sc}$.}
\label{fig:setup_ideal}
\end{figure}

\section{Ideal measurement scheme}
We consider a laser beam, linearly polarized along the $x$ direction and propagating along the optical $z$ axis, which is strongly focused, as shown in Fig.~\ref{fig:setup_ideal}. The  focus defines the origin of the coordinate system. We furthermore assume an isotropic dipolar scatterer with polarizability $\alpha$ located at position $\vect{r}_0 = (x_0, y_0, z_0)$, whose distance to the focus is much less than a wavelength. At the scatterer position, the electric field reads
\begin{equation}
 \mathbf{E}_{\rm foc}(\vect{r}_0) = E_0 \mathbf{n}_x \exp(\imu Akz_0),
\end{equation}
which resembles a plane wave travelling into the positive $z$ direction (see Appendix~\ref{app:focal_field} for a derivation of the focal field). Here, $\mathbf{n}_x$ is the unit vector along $x$, $E_0$ is the field amplitude at the focus, and $k$ is the wavenumber. The geometric factor $0.64<A\le1$ is a result of the Gouy phase shift in a focused beam and increases the effective wavelength close to the focus. We derive an analytical expression for $A$ in a strongly focused field in Appendix~\ref{app:focal_field}. In a mildly focused field, described as a Gaussian beam with Rayleigh range $z_R$, we find $A=1-(kz_R)^{-1}$~\cite{Novotny2012,Dawson2019}.

The dipolar scatterer acquires the dipole moment $\mathbf{p} = \alpha \mathbf{E}_{\rm foc}(\vect{r}_0)$, and radiates the scattered field $\mathbf{E}_{\rm sc}(\mathbf{r})$. At an observation point $\vect{r}$ much farther from the scatterer than a wavelength, we can write the scattered field in the Fraunhofer approximation as
\begin{equation}\label{eq:dipole_scattered}
    \mathbf{E}_{\rm sc}(\mathbf{r}) = \mathbf{E}_{\rm dip}(\mathbf{r}) \exp\left[-\imu k \cdot \left(\mathbf{r_0} \cdot \mathbf{n}_r - A z_0 \right)\right],
\end{equation}
where $\vect{n}_r$ is the unit vector in the radial direction and $\vect{E}_\mathrm{dip}(\vect{r})$ is the far-field generated at the observation point $\vect{r}$ by an $x$ oriented dipole located at the origin.
Importantly, the scatterer's position $\mathbf{r}_0$ is contained in the phase of the scattered field, which consists of two terms. The first term $-k\mathbf{r}_0 \cdot \mathbf{n}_r$ describes the phase generated by the displacement of the dipole relative to the origin. The second term $A k z_0$ stems from the fact that the dipole is driven by a travelling wave which acquires a phase shift during propagation. In the following, we use a spherical coordinate system, where the angle $\theta$ denotes the polar angle relative to the $z$ axis, and $\phi$ the azimuthal angle relative to $x$. Furthermore, we introduce the differential power $\diff p_{\rm dip}$ scattered by an $x$ polarized dipole into the solid angle $\diff\Omega = \sin(\theta)\diff\theta\diff\phi$ 
    \begin{equation}\label{eq:pdip}
        \diff p_{\rm dip} = \frac{3}{8\pi} P_{\rm dip}\left[ 1-\sin(\theta)^2 \cos(\phi)^2 \right]~\diff\Omega,
    \end{equation}
with the total scattered power $P_{\rm dip}$.

\subsection{Measurement backaction}\label{sec:msmt_backaction}
In this subsection, we analyze the measurement backaction arising from the interaction of the scatterer with the electromagnetic field. This backaction takes the form of a recoil force, which can be interpreted as an inevitable consequence of the fact that the scattered field contains information about the scatterer's position. 
The measurement-backaction force along a certain direction $(x,y,z)$ can be quantified by its power spectral density~\footnote{We define our power spectral densities according to $\Braket{y^2} = \int_{-\infty}^{\infty} {\rm d}\Omega ~ S_{yy}(\Omega)$}.
Along the transverse directions $x$ and $y$, the spectral densities of this measurement backaction read~\cite{Jain2016, Schottkey1918}
\begin{subequations}\label{eq:Sff}
    \begin{align}
        \label{eq:Sff_x} S_{\rm ba}^x & =\frac{1}{5} \frac{\hbar k}{2\pi c} P_{\rm dip}, \\
        \label{eq:Sff_y} S_{\rm ba}^y & =\frac{2}{5} \frac{\hbar k}{2\pi c} P_{\rm dip}.
    \intertext{Note that along the $x$ and $y$ axes, the result for the measurement backaction for a (passive) dipole scattering a power $P_\mathrm{dip}$ equals the result for an active dipolar source radiating the same power. In contrast to an active source, however, the fluctuating force acting on a scatterer along the $z$ axis has two contributions. First, there is the contribution that equals $S_\text{ba}^{y}$, which arises both for a passive scatterer and an active dipolar source. However, for a passive scatterer, a second term arises, which stems from the fluctuations of the radiation pressure along the propagation direction of the beam. In close vicinity of the focus, for a lossless dipolar scatterer, this radiation pressure force reads $F_{\rm rp}^z = A P_{\rm dip}/c$~\cite{Novotny2012}.  Summing both contributions, we find the total measurement backaction along the $z$ axis
    }
        \label{eq:Sff_z} S_{\rm ba}^z & =\left(\frac{2}{5} + A^2 \right) \frac{\hbar k}{2\pi c} P_{\rm dip}.
    \end{align}
\end{subequations}
We provide a derivation of Eqs.~\eqref{eq:Sff} in Appendix~\ref{app:deriv_backaction} and note that our results agree with a full quantum calculation~\cite{Gordon1980}.

\subsection{Measurement imprecision}\label{sec:measurement_imprecision}
Having dealt with the measurement backaction, we now turn to the measurement imprecision associated with locating a point scatterer in a focused light field. Since the scatterer's position $\mathbf{r}_0$ is encoded solely in the phase of the scattered light according to Eq.~\eqref{eq:dipole_scattered}, we make use of a homodyne measurement, where we superpose the scattered light at the detector position $\mathbf{r}$ with a strong local oscillator field. Note that we do \emph{not} use the trapping field as a reference field here (as is done in typical experimental schemes discussed in Sec.~\ref{sec:real_det_system}), which is always possible by introducing a sufficiently strong additional reference. For optimal interference, we choose the (local) polarization of the reference to equal that of the scattered light. Thus, we construct an \emph{ideal} reference field 
    \begin{equation}\label{eq:referenceField}
        \mathbf{E}_{\rm ref}(\mathbf{r}) = -\imu\gamma \mathbf{E}_{\rm dip}(\mathbf{r})   
    \end{equation}
with $\gamma \gg 1$ (such that the reference field is much stronger than the scattered field) and let it interfere with the dipole's scattered light such that the field at position $\mathbf{r}$ is $\mathbf{E}_{\rm ref}(\mathbf{r}) + \mathbf{E}_{\rm sc}(\mathbf{r})$. For small displacements $\vect{r}_0$, a detector positioned at $\mathbf{r}$ covering the differential solid angle $\diff\Omega$ measures the power
        \begin{equation}\label{eq:Pdet_ideal}
            \diff p_{\rm det}(\theta, \phi) = \left[\gamma^2 +2\gamma k \left(\mathbf{r}_0 \cdot \mathbf{n}_r - A z_0 \right)\right]~\diff p_{\rm dip},
        \end{equation}
where we have retained only the term linear in the scattered field (which is much weaker than the reference field). The first term of Eq.~\eqref{eq:Pdet_ideal} accounts for the power of the reference and the second term for the interference between reference and scattered fields. Being linear in $\vect{r}_0$, the interference term represents a measure of position.
Assuming shot noise as the dominating noise source, the power spectral density of the imprecision noise associated with a differential detector under direction $(\theta,\phi)$ is (see Appendix~\ref{app:deriv_imprecision} for detailed derivation)
\begin{subequations}\label{eq:DeltaS_imp_all}
        \begin{align}
            s_\mathrm{imp}^{x}(\theta,\phi) &= \frac{\hbar c}{8\pi k \sin(\theta)^2 \cos(\phi)^2    ~  \diff p_{\rm dip}}, \label{eq:diffSimpX}\\
            s_\mathrm{imp}^{y}(\theta,\phi) &= \frac{\hbar c}{8\pi k \sin(\theta)^2 \sin(\phi)^2    ~  \diff p_{\rm dip}}  ,\label{eq:diffSimpY}\\
            s_\mathrm{imp}^{z}(\theta,\phi) &= \frac{\hbar c}{8\pi k [\cos(\theta)-A]^2    ~  \diff p_{\rm dip}}\label{eq:diffSimpZ}.
        \end{align}
\end{subequations}
%
Next, we fill up the entire sphere surrounding the scatterer with differential detectors. Importantly, Eqs.~\eqref{eq:DeltaS_imp_all} show that the measurement imprecision depends on the position $(\theta,\phi)$ of the differential detector and is not uniformly distributed. 
Accordingly, to minimize the total imprecision, we need to weight each measurement by its inverse imprecision before averaging~\cite{Strutz2010}.
To obtain the total imprecision, we exploit the fact that its inverse $1/S_\text{imp}$ is given by the integral over the inverse imprecisions contributed by the differential detectors $1/[s_\text{imp}(\theta,\phi)]$, such that we find (the calculation is detailed in Appendix~\ref{app:deriv_imprecision})
    \begin{subequations}\label{eq:Simp_all}
    \begin{flalign}
    \label{eq:Sxx} 
        S_\mathrm{imp}^{x} & = 5\frac{\hbar c}{8\pi k} \frac{1}{P_{\rm dip}},\\
    \label{eq:Syy}
        S_\mathrm{imp}^{y} & = \frac{5}{2} \frac{\hbar c}{8\pi k}  \frac{1}{P_{\rm dip}},\\
    \label{eq:Szz}
        S_\mathrm{imp}^{z} & = \frac{1}{\frac{2}{5} +A^2} \frac{\hbar c}{8\pi k} \frac{1}{P_{\rm dip}}.
    \end{flalign}
    \end{subequations}
    
\begin{figure}
\includegraphics[width=\columnwidth]{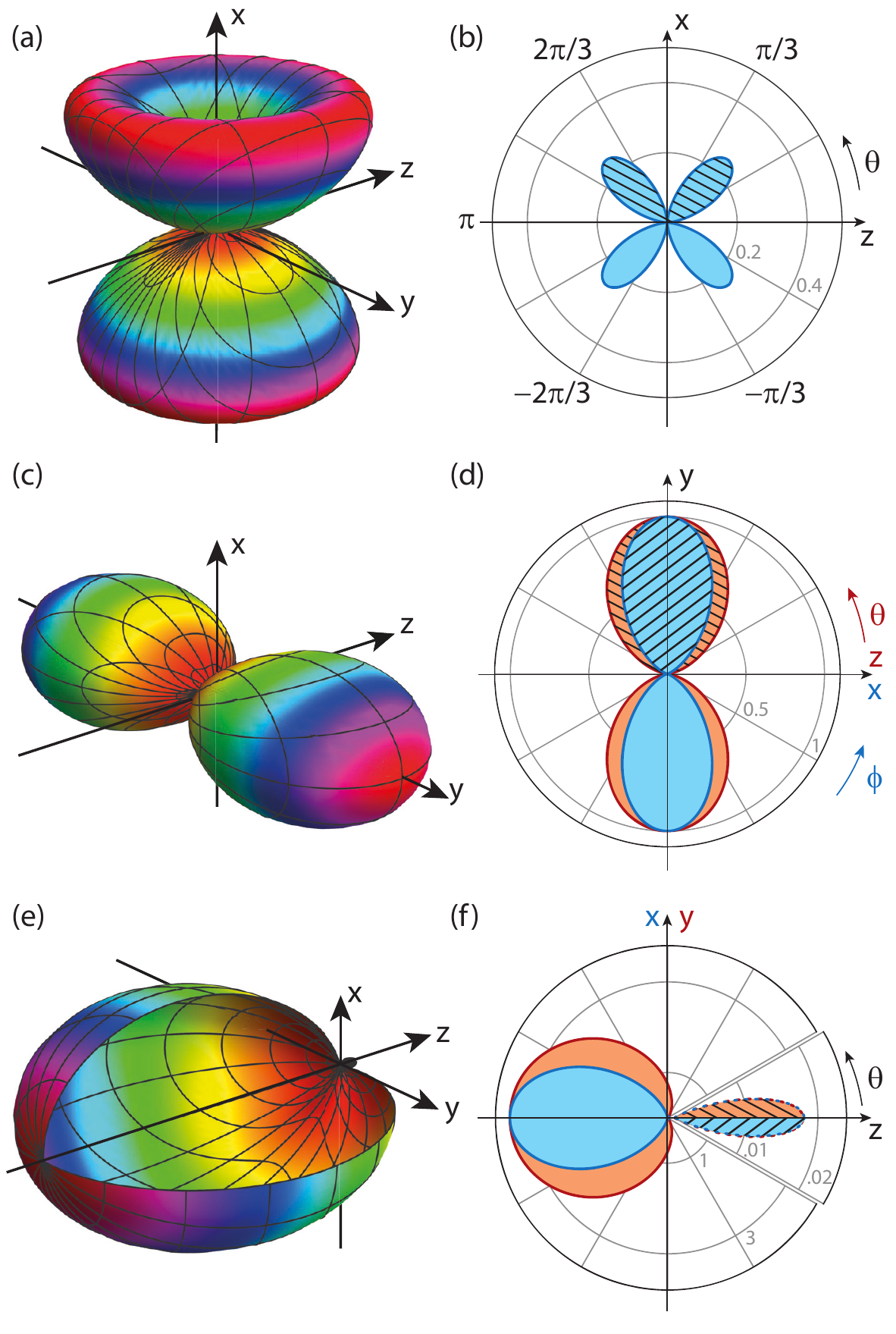}
\caption{Information radiation patterns. (a)~Contour plot of the information radiated into a unit solid angle. We plot the quantity $\mathcal{I}_x(\theta,\phi)$ as the radial distance of the contour to the origin. (b) Cross section through $\mathcal{I}_x$ in the plane $xz$. The cross-hatched region indicates where a displacement $x_0>0$ gives rise to a positive detector signal $\diff p_\text{det}(\theta,\phi)$. (c)~Contour plot of $\mathcal{I}_y$. (d)~Cross sections of $\mathcal{I}_y$ in the plane $xy$ (blue) and $yz$ (red). (e)~Contour plot of $\mathcal{I}_z$. Note that the information is mostly radiated in the negative $z$ direction. (f)~Cross sections of $\mathcal{I}_z$ in the $xz$ plane (blue) and in the $yz$ plane (red). Note the different radial scale in the range $-\pi/6\le\theta\le\pi/6$. $A$ was fixed to $\cos(\pi/6) = 0.866$.}
\label{fig:informationDirection}
\end{figure}

\subsection{Discussion of ideal measurement scheme}\label{sec:discussion_ideal_measurement}
By comparing Eqs.~\eqref{eq:Simp_all} with Eqs.~\eqref{eq:Sff}, we find that the imprecision-backaction product $S_\mathrm{imp}^j S_{\rm ba}^j = [\hbar/(4\pi)]^2$ fulfills the Heisenberg uncertainty relation with equality for all three axes $j\in\{x,y,z\}$~\cite{Clerk2010}. Accordingly, our measurement scheme decodes the scatterer's position in an optimal way along all three axes simultaneously. 
It is instructive to consider the angular dependence of the contributions to the measurement imprecision in Eqs.~\eqref{eq:DeltaS_imp_all}. 
To this end, we inspect their (normalized) inverse 
$\mathcal{I}_j(\theta,\phi)=S_\text{imp}^j / s_\text{imp}^j(\theta,\phi)$. 
This quantity resembles the angular information density about the scatterer's position along the axis $j$. As an example, let us consider motion along the $x$ axis, shown in Fig.~\ref{fig:informationDirection}(a). We plot $\mathcal{I}_x(\theta,\phi)$ such that its value is encoded as the radial distance of the contour to the origin. We observe that the information content $\mathcal{I}_x$ vanishes in the plane $x=0$. This means that a detector located anywhere in this plane cannot extract any information about the scatterer's position along $x$. This observation makes sense, since any displacement along $x$ (to linear order) has no influence on the phase of the field scattered in the plane $x=0$. Furthermore, also a detector located on the (positive or negative) $x$ axis cannot infer any information about the motion along $x$. This observation might be surprising at first sight, since the phase shift of the scattered signal along this direction should be most sensitive to the scatterer's position along $x$. However, a linearly polarized dipole radiates no far-field along its axis and the measurement signal vanishes along the $x$ axis. In Fig.~\ref{fig:informationDirection}(b), we show a cross section of $\mathcal{I}_x$ in a plane containing the $x$ axis (which is an axis of symmetry for $\mathcal{I}_x$). We have cross-hatched the region where the signal $\diff p_\mathrm{det}(\theta,\phi)$ is positive for a positive displacement $x_0$ of the scatterer, which is the case in the half-space $x>0$. 
In Fig.~\ref{fig:informationDirection}(c), we show $\mathcal{I}_y(\theta,\phi)$. We see that most information about the scatterer's position along $y$ is radiated along the $y$ axis. This observation makes sense, since both the radiation pattern of the dipolar scatterer and the dependence of the phase of the scattered field on the position along the $y$ axis reach a maximum along that direction. In Fig.~\ref{fig:informationDirection}(d), we show a cross section of $\mathcal{I}_y$ in the $yz$ plane (red) and in the $xy$ plane (blue). The signal $\diff p_\mathrm{det}(\theta,\phi)$ is positive for positive $y_0$ in the half-space $y>0$.

The quantity $\mathcal{I}_z(\theta,\phi)$ shown in Fig.~\ref{fig:informationDirection}(c), is particularly interesting. In contrast to $\mathcal{I}_x$ ($\mathcal{I}_y$), which bears a symmetry relative to the plane $x=0$ ($y=0$), $\mathcal{I}_z$ has no symmetry relative to $z=0$. This symmetry is broken by the propagating nature of the beam illuminating the scatterer. It turns out that more than 90\% of the entire information about the position along the $z$ axis is contained in the field scattered in the backward direction (half-space $z<0$). This observation can be intuitively understood in the limiting case of a plane wave illuminating the scatterer ($A=1$). 
In this case, the phase of the field scattered in the forward direction on the optical axis is independent of the scatterer's position along the $z$ axis. 

Let us recap at this point the essential features of the optimal measurement scheme discussed thus far. The first feature is the optimal reference field in Eq.~\eqref{eq:referenceField}. This position-dependent field has to locally match the polarization of the field radiated by the scatterer. Second, the optimal reference field has to be phase-shifted by $\pi/2$ relative to the scattered field. Finally, the differential detector signal collected on a detector under the direction $(\theta,\phi)$ has to be appropriately weighted according to its inverse imprecision noise as given by Eqs.~\eqref{eq:DeltaS_imp_all} to obtain the optimal measurement of the scatterer's position. The total measurement imprecision of this scheme, given by Eqs.~\eqref{eq:Simp_all}, multiplies with the measurement backaction given by Eqs.~\eqref{eq:Sff} to fulfill the Heisenberg uncertainty relation with equality in each direction $(x,y,z)$.

\section{Realistic detection system} \label{sec:real_det_system}
Thus far, we have analyzed the problem of detecting the position of a dipolar scatterer and described an ideal measurement scheme that allows for a Heisenberg-limited measurement of the scatterer's position in three dimensions. 
Two experimental difficulties make our ideal measurement scheme impractical. First, it is challenging to generate a reference field whose phase and polarization match those of a dipolar field, as required by Eq.~\eqref{eq:referenceField}. Reference fields typically available in a laboratory setting are Gaussian beams with uniform polarization. Second, this ideal measurement scheme requires a distribution of infinitesimal detectors spanning the full $4\pi$ of solid angle, where the signal from each detector is individually weighted according to its imprecision. In contrast, in practice one typically uses a simple four-quadrant detector~\cite{Gittes1998}. 

\begin{figure*}
\includegraphics[width=\textwidth]{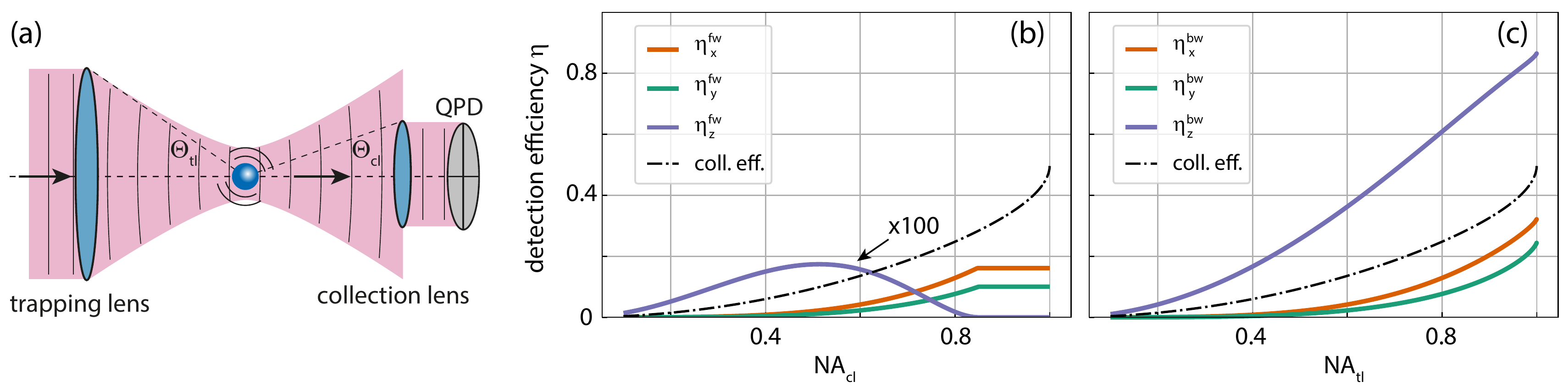}
\caption{
(a)~Laboratory detection system. A trapping lens with numerical aperture $\NA_\mathrm{tl}=\sin(\Theta_\mathrm{tl})$ focuses an $x$ polarized plane wave. On the opposite side, the fields are collimated by a collection lens with numerical aperture $\NA_\mathrm{cl}=\sin(\Theta_\mathrm{cl})$. A particle close to the focal point scatters the focused field. 
(b)~Detection efficiency $\eta^\text{fw}$ in forward detection for a trapping lens with $\NA_\text{tl}=0.85$. We plot the detection efficiencies for motion in the focal plane $\eta_x^\mathrm{fw}(\NA_\mathrm{cl})$ (red solid),  $\eta_y^\mathrm{fw}(\NA_\mathrm{cl})$ (green solid) as a function of $\NA_\text{cl}$. The detection efficiency along the optical axis $\eta_z^\mathrm{fw}(\NA_\mathrm{cl})$ (blue solid, multiplied by factor 100), vanishes when $\NA_\mathrm{tl}=\NA_\mathrm{cl}$. In the range $\NA_\mathrm{cl}>\NA_\mathrm{tl}$, the detection efficiencies stay constant, since no reference field is available in that range. We also plot the fraction of the scattered power which is collected by the optics (black dash-dotted). 
(c)~Detection efficiency $\eta^\text{bw}$ in backward detection. In the transverse directions, $\eta^\text{bw}$ is identical to the case of forward detection shown in (b). However, most information about the motion along $z$ is encoded in the backscattered field, such that $\eta_z^\mathrm{bw}$ (blue dashed) reaches values exceeding 0.6 for realistic trapping lenses with $\NA_\mathrm{tl}>0.8$. }
\label{fig:referenceSpheres_det_eff}
\end{figure*}

\subsection{Forward detection}
In this section, we consider the performance of the most commonly used detection system in optical trapping experiments, which relies on a standard (four-quadrant) split detection scheme in forward direction~\cite{Gittes1998}. The situation under consideration is sketched in Fig.~\ref{fig:referenceSpheres_det_eff}(a). A first lens (termed the `trapping lens') focuses an $x$ polarized plane wave (corresponding to a strongly overfilled objective) such that the focal point coincides with the origin. The optical axis is along $z$, and a second lens (termed the `collection lens') recollimates the trapping beam. 
A dipolar scatterer is located close to the origin and generates the scattered field $\mathbf{E}_\text{sc}(\vect{r})$ given by Eq.~\eqref{eq:dipole_scattered}. For detection in forward direction, the reference field Eq.~\eqref{eq:referenceField} has to be replaced by the field of the trapping beam arriving on the detector. 

As detailed in Appendix~\ref{app:deriv_realistic_detection}, and in analogy to Section~\ref{sec:measurement_imprecision}, we calculate the measurement imprecision $S_\text{imp}^{j,\text{fw}}$ of this forward detection scheme for all three axes $j\in\{x,y,z\}$. 
In order to compare our results for forward detection with the ideal case discussed in Section~\ref{sec:measurement_imprecision}, we define the detection efficiency $\eta_j^\text{fw}=S_\text{imp}^{j}/S_\text{imp}^{j,\text{fw}}$ for the axis $j$ as the ratio of the result for forward scattering $S_\text{imp}^{j,\text{fw}}$ and the measurement imprecision at the Heisenberg limit given by Eqs.~\eqref{eq:Simp_all}. Thus, the detection efficiency is a measure for how close to the Heisenberg limit a detection system operates. Note that absorption losses or a limited quantum efficiency of the detector further decrease the detection efficiency.

In Fig.~\ref{fig:referenceSpheres_det_eff}(b), we plot the detection efficiencies $\eta_x^\mathrm{fw}$ (red), $\eta_y^\mathrm{fw}$ (green), and $\eta_z^\mathrm{fw}$ (blue) in forward-scattering as a function of the numerical aperture of the collection lens $\NA_\text{cl}$ for a numerical aperture of the trapping lens $\NA_\text{tl}=0.85$. 
As expected, a larger $\NA_\mathrm{cl}$ generally leads to higher detection efficiency for all three axes. 
However, the detection efficiency for the $z$ axis $\eta_z^\mathrm{fw}$ shows a remarkable feature. It turns out that $\eta_z^\mathrm{fw}$ vanishes for a symmetric setup, i.e., when the numerical aperture of the trapping lens equals that of the collection lens. 
We discuss this feature quantitatively in Appendix~\ref{app:deriv_realistic_detection}. It can be understood qualitatively by close inspection of Fig.~\ref{fig:informationDirection}(f) from the discussion of the ideal measurement scheme. As indicated by the cross-hatched regions in the polar plot, the signal changes sign in the half-space $z>0$. In the specific case of a symmetric setup ($\NA_\mathrm{tl}=\NA_\mathrm{cl}$), the integration over $\theta$ is truncated such that the result strictly vanishes.

Let us apply our insights to realistic experimental conditions. For typical values of $\NA_\mathrm{cl}=0.7$, the detection efficiency of the transverse modes in forward direction is around $\eta_x^\mathrm{fw}\sim\eta_y^\mathrm{fw}\sim 0.1$, while for the longitudinal mode it is about two orders of magnitude smaller ($\eta_z^\mathrm{fw}\sim 0.001$).

\subsection{Backward detection}
We now turn to detection in backward direction. Here, the backscattered light is collected by the trapping objective and then interfered with an external reference field~\cite{Vovrosh2017}. In contrast to forward scattering, where the reference beam is naturally phase locked with the right phase shift (due to the common-path arrangement together with the Gouy phase shift), backward scattering is technically more involved, since the phase shift of the reference beam relative to the scattering signal has to be actively stabilized to a value of $\pi/2$. 
To compare with forward scattering, we consider a reference field that has the same spatial distribution as the trapping beam (a truncated plane wave).

In Appendix~\ref{app:BackwardDetection}, we derive expressions for the detection efficiencies $\eta_j^\text{bw}$ in backward direction, which are plotted in Fig.~\ref{fig:referenceSpheres_det_eff}(c) as a function of the numerical aperture of the trapping lens $\NA_\mathrm{tl}$.
We find that the detection efficiencies for motion in $x$ and $y$ directions are the same for forward and backward detection, i.e.,  $\eta_x^\mathrm{bw}=\eta_x^\mathrm{fw}$ and $\eta_y^\mathrm{bw}=\eta_y^\mathrm{fw}$. This result is expected, since the information about motion along $x$ and $y$ is radiated symmetrically in the forward and in the backward direction [compare  Figs.~\ref{fig:informationDirection}(a,c)]. On the other hand, for motion along the $z$ direction, we find that the detection efficiency is much higher in backward direction than in forward direction. This result can be anticipated from the distribution of radiated information content shown in Figs.~\ref{fig:informationDirection}(e,f). For a typical value of the numerical aperture of the trapping lens $\NA_\mathrm{tl}=0.8$, we find the efficiency to be as high as $\eta_z^\mathrm{bw}=0.6$.

\subsection{Discussion of real-world measurement schemes}
Let us recap the most important insights gained from our analysis. Clearly, forward and backward detection using quadrant detectors fall short of reaching the Heisenberg limit of maximum detection efficiency $\eta=1$, where the imprecision-backaction product has its minimum. Several factors contribute to this imperfection. First, the numerical aperture collecting the light scattered by the dipole is finite and, as a result, part of the information about the dipole's position is not collected. We plot the fraction of collected power as a function of numerical aperture in Figs.~\ref{fig:referenceSpheres_det_eff}(b,c) as the black dash-dotted lines. Importantly, not every collected photon carries the same amount of information. For example, motion along the $z$ axis is predominantly encoded in the field scattered in backward direction, allowing for a large detection efficiency for the $z$ motion in backscattering. Another factor limiting the detection efficiency is the imperfect overlap of the reference field (with homogeneous polarization) with the field scattered by the dipole (whose polarization varies spatially). Finally, an ideal detection system must not only collect the measurement signal across the full solid angle surrounding the scatterer, but must also weight the individual contributions according to their information content as given by the measurement imprecision. Clearly, a quadrant detector offers very limited capability to perform this weighting procedure.
Consider, for example, detection of the $y$ motion. It is clear from Fig.~\ref{fig:informationDirection} that practically no information is contained in the signal striking the detector close to the $z$ axis. Nevertheless, a standard detector will sum the shot-noise contribution generated in this region by the reference field and add it to the output signal. 
A possible alternative to a spatially resolving detector would be a reference field with an appropriately shaped spatial intensity distribution. Such a field distribution could, for example, be generated using a spatial light modulator.

Finally, let us consider the repercussions of our findings for active feedback cooling of a levitated nanoparticle's motion. Considering the finite transmissivity of optical components, the finite quantum efficiency of photodetectors, and our finding that the detection efficiency for the motion along the optical axis in backward scattering can reach $\eta_z^\mathrm{bw}\sim0.8$, we conclude that a total efficiency of 0.35 appears well within reach.
Adding the fact that at sufficiently low pressures, the reheating of a levitated particle is dominated by measurement backaction~\cite{Jain2016}, active feedback by means of cold damping~\cite{Tebbenjohanns2019,Conangla2019} should be able to cool a levitated nanoparticle in a free-space configuration with only a single laser beam to mean phonon occupation numbers as low as $n = [\eta^{-1/2} - 1]/2 = 0.35$ along the optical axis, and thus to the quantum ground state of motion.

\section{Conclusions} 
We have theoretically analyzed the problem of determining the position of a dipolar scatterer in a focused field. We have proposed an ideal detection scheme locating the scatterer in three dimensions at the Heisenberg limit of the imprecision-backaction product. Furthermore, we have analyzed configurations commonly used in experiments and derived their measurement efficiencies. We have found that for realistic experimental setups, the detection efficiencies for motion transverse to the optical axis are limited to $\eta_{x,y}\sim0.1$. On the other hand, our analysis shows that the motion along the optical axis is most efficiently detected in backscattering, where the detection efficiency of the longitudinal motion can be as high as 80\%, such that ground-state cooling of a levitated particle in a single beam optical trap should be feasible.

\begin{acknowledgments}
This research was supported by ERC-QMES (Grant No. 338763), the NCCR-QSIT program (Grant No. 51NF40-160591), and SNF (Grant No. 200021L-169319/1). We thank V.~Jain, F.~van der Laan, A.~Militaru, R.~Reimann,  and D.~Windey for valuable input and discussions.
\end{acknowledgments}

\bibliography{Literature_Tebbenjohanns_Efficiency}

\appendix
\section{Effective wavelength of focused field}\label{app:focal_field}
In this Appendix, we show that a strongly focused field in the focal region to first order appears as a plane wave propagating along the optical axis with an effective wavelength determined by the numerical aperture of the focusing lens.
We start with the focal field generated by a highly overfilled objective, which can be written analytically in cylindrical coordinates as~\cite{Novotny2012}
\begin{equation}
\label{eq:focal_field_exact}
\mathbf{E}_{\rm foc} \propto \begin{pmatrix} I_{00} + I_{02} \cos (2\phi) \\ I_{02}\sin (2\phi) \\ -2\imu I_{01} \sin(\phi) \end{pmatrix},
\end{equation}
where the incoming light is polarized along $x$. The integrals $I_{00}$, $I_{01}$, and $I_{02}$ depend on coordinates $\rho$ and $z$
\begin{subequations}
\label{eq:focal_field_integrals}
\begin{align}
I_{00} =& \int_0^{\Theta_{\rm tl}} \diff\theta~ \sqrt{c} s (1 + c) J_0(k\rho s) \mathrm{e}^{\imu kz c} \\
I_{01} =& \int_0^{\Theta_{\rm tl}} \diff\theta~ \sqrt{c} s^2 J_1(k\rho s) \mathrm{e}^{\imu kz c} \\
I_{02} =& \int_0^{\Theta_{\rm tl}} \diff\theta~ \sqrt{c} s (1-c) J_2(k\rho s) \mathrm{e}^{\imu kz c} ,
\end{align}
\end{subequations}
where $c=\cos(\theta)$ and $s=\sin(\theta)$. Furthermore, $J_n$ are the Bessel functions of the first kind for $n\in\mathbbm{N}_0$, which we expand to first order as $J_0(x) = 1$, $J_1(x) = x/2$, and $J_2(x) = 0$. The $x$ component of $\mathbf{E}_{\rm foc}$ to first order reads $C + \imu kz D \approx C \exp(\imu kz D/C)$ where $C$ and $D$ are integrals over $\theta$ which are independent of any coordinates.
The electric field component along $y$ vanishes to first order. The phase of the $z$ component of the field is constant to first order, but its amplitude is linear in the transverse directions and vanishes at the origin. Importantly, the $z$ component of the field is $\pi/2$ out of phase with the $x$ polarization such that it appears only to second order in standard homodyne detection schemes. We hence conclude that the focal field in close vicinity to the focus can be approximated as an $x$ polarized plane wave traveling in the positive $z$ direction according to $\mathbf{E}_{\rm foc} = E_0 \mathbf{n}_x \exp (\imu Akz)$, with 
\begin{equation}
\label{eq:A}
A = \frac{\int_0^{\Theta_{\rm tl}}\diff\theta ~ s \sqrt{c} (1+c)c}{\int_0^{\Theta_{\rm tl}}\diff\theta ~ s \sqrt{c} (1+c)}.
\end{equation}
See App.~\ref{app:analytical_solutions} for an analytical expression for $A$.

\section{Derivation of measurement backaction}
\label{app:deriv_backaction}
The differential power $\diff p_\text{dip}$ radiated by an $x$ polarized dipolar scatterer into solid angle $\diff \Omega = \sin(\theta)~ \diff\theta ~\diff\phi$ is given by Eq.~\eqref{eq:pdip}. This power exerts a radiation pressure force $\diff F_\text{rp}^x = - (\vect{n}_x \cdot \vect{n}_r) \diff p_\text{dip} / c$ on the scatterer along the $x$ direction. Assuming shot noise to dominate the fluctuations of the radiated power, we find for the power spectral density of the radiated power along direction $\vect{n}_r$ 
\begin{equation}
    \diff s_\text{pp}(\theta, \phi) = \frac{\hbar k c}{2\pi} \diff p_\text{dip}.
\end{equation}
Due to these fluctuations, the radiation pressure force along $\vect{n}_x$ fluctuates with power spectral density $\diff s_\text{ba}^x = (\vect{n}_x \cdot \vect{n}_r)^2 ~\diff s_\text{pp}(\theta,\phi) /c^2$. We integrate this differential contribution to the measurement backaction over the unit-sphere to find the backaction noise spectral density along the $x$ direction
\begin{equation}
    \label{eq:Sff_x_deriv} S_{\rm ba}^x = \int~\diff s_\text{ba}^x(\theta,\phi) = \frac{1}{5} \frac{\hbar k}{2\pi c} P_{\rm dip}.
\end{equation}
By analogous derivations, we find the values for $S_\text{ba}^y$ and $S_\text{ba}^z$ as given by Eqs.~\eqref{eq:Sff}. As described in the main text, along the $z$ direction, $S_\text{ba}^z$ needs to be amended by an additional contribution for a scatterer polarized by a travelling wave, which exerts a radiation pressure force along its propagation direction.
While our derivation rests on a semiclassical treatment, our results match a full quantum derivation for a two-level system in the classical limit~\cite{Gordon1980}.

\section{Derivation of optimal measurement imprecision}\label{app:deriv_imprecision}
In this Appendix, we derive the measurement imprecision of our ideal measurement scheme.
Equation~\eqref{eq:Pdet_ideal} in the main text is the differential power impinging on a detector covering a solid angle $\diff\Omega$ and located at $(\theta, \phi)$. Let us temporarily assume that the scatterer is only displaced along the $x$ axis, such that $y_0=z_0 = 0$. Then, the locally measured power
\begin{equation}\label{eq:pdet_ideal_x}
\diff p_\text{det}(\theta, \phi) = [\gamma^2 + 2\gamma k \sin(\theta)\cos(\phi) x_0]\diff p_\text{dip}
\end{equation}
linearly depends on position $x_0$ and therefore is a measure for the scatterer's position $x_0$. The first term contributing to $\diff p_\mathrm{det}$ is independent of $x_0$, but dominates the fluctuations of the measurement through photon shot noise.
The power spectral density of the fluctuations is
\begin{equation}\label{eq:dspp_ideal}
    \diff s_\text{pp}^\text{det}(\theta, \phi) = \frac{\hbar k c}{2\pi} \gamma^2 \diff p_\text{dip}.
\end{equation}
In order to extract the position $x_0$ from the differential detector signal Eq.~\eqref{eq:pdet_ideal_x}, it needs to be divided by the pre-factor $\diff\beta(\theta, \phi) = 2\gamma k \sin(\theta) \cos(\phi) \diff p_\text{dip}$. Accordingly, we translate the fluctuations given in Eq.~\eqref{eq:dspp_ideal} to fluctuations of the position as
\begin{equation}
    s_\text{imp}^x(\theta, \phi) = \frac{\diff s_\text{pp}^\text{det}(\theta, \phi)}{\diff\beta(\theta, \phi)^2}. 
\end{equation}
In analogy, we derive $s_\text{imp}^y(\theta, \phi)$ and $s_\text{imp}^z(\theta, \phi)$ given in Eqs.~\eqref{eq:DeltaS_imp_all}. Recall that $s_\text{imp}^j(\theta, \phi)$ for $j\in\{x,y,z\}$ are the power spectral densities of the measurement imprecision associated with a differential detector located at $(\theta, \phi)$.  Importantly, we find $s_\text{imp}^j(\theta, \phi)\propto 1/(\diff\Omega)$, meaning that as the solid angle $\diff\Omega$ goes to zero, the imprecision noise of the detector diverges to infinity. This intuitively makes sense, since the signal vanishes together with the detector area.

As mentioned in the main text, we perform inverse-variance weighting in order to minimize the total imprecision when the signals from all detectors covering the unit-sphere are combined. The local reading of $x_0$ is hence weighted with the inverse of $s_\mathrm{imp}^{x}(\theta, \phi)$ before averaging over the unit sphere. The total imprecision then turns out to be~\cite{Strutz2010}
\begin{equation}
    S_\mathrm{imp}^{x} = \left[\int \frac{\diff\beta(\theta, \phi)^2}{\diff s_\text{pp}^\text{det}(\theta, \phi)}\right]^{-1}=  5\frac{\hbar c}{8\pi k} \frac{1}{P_{\rm dip}},
\end{equation}
where the integral runs over the full unit-sphere. The results for all three axes are given in Eqs.~\eqref{eq:Simp_all}.
        
Finally, let us drop our assumption that the scatterer is only displaced along one axis and allow for the position $\vect{r}_0$ to have three non-zero components. In this case, $\diff p_\text{det}$ depends on a linear superposition of $x_0$, $y_0$, and $z_0$ and, therefore, also the local reading for the position along $x$ 
    \begin{equation}\label{eq:diff_pos_signal}
    \begin{split}
        \tilde{x}_0(\theta,\phi) &= \frac{\diff p_\text{det}(\theta,\phi) - \gamma^2 \diff p_\text{dip}}{\diff\beta(\theta,\phi)} \\
        &= x_0 + y_0 \tan(\phi) + z_0\frac{\cos(\theta) - A}{\sin(\theta) \cos(\phi)}
    \end{split}
    \end{equation}
includes contributions by $y_0$ and $z_0$. Nevertheless, when combining the information from all differential detectors correctly, following the procedure of inverse variance weighting, we indeed find
        \begin{equation}       \label{eq:extractionRecipes_all}
            x_0= S_\mathrm{imp}^{x} \int \frac{\diff\beta(\theta, \phi)^2}{\diff s_\text{pp}^\text{det}(\theta, \phi)}~ \tilde{x}_0(\theta,\phi),
        \end{equation} 
meaning that the contributions of $y_0$ and $z_0$ cancel. 
Analogously, $y_0$ and $z_0$ can be extracted from the measurements $\diff p_\mathrm{det}$ using
    \begin{subequations}\label{eq:diff_pos_signal_yz}
    \begin{align}
        \tilde{y}_0(\theta,\phi) =& \frac{\diff p_\text{det} - \gamma^2\diff p_\text{dip}}{2\gamma k \sin(\theta) \sin(\phi) \,\diff p_\mathrm{dip}},\\
        \tilde{z}_0(\theta,\phi) =& \frac{\diff p_\text{det} - \gamma^2\diff p_\text{dip}}{2\gamma k [\cos(\theta)-A] \,\diff p_\mathrm{dip}},
    \end{align}
    \end{subequations}
together with an appropriately adjusted version of Eq.~\eqref{eq:extractionRecipes_all}.

\section{Derivation of realistic measurement imprecision}\label{app:deriv_realistic_detection}
\begin{figure}
\includegraphics[width=\columnwidth]{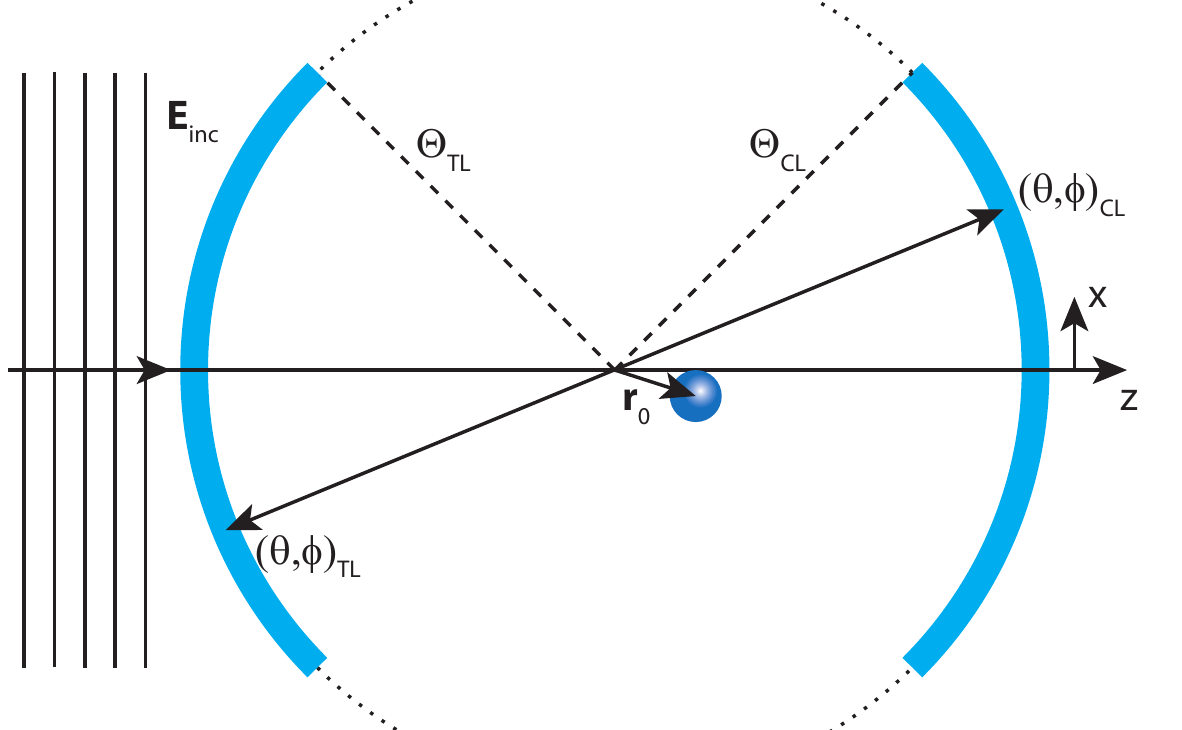}
\caption{Laboratory detection system. A trapping lens with numerical aperture $\NA_\mathrm{tl}=\sin(\Theta_\mathrm{tl})$ focuses an $x$ polarized plane wave. On the opposite side, the fields are collimated by a collection lens with numerical aperture $\NA_\mathrm{cl}=\sin(\Theta_\mathrm{cl})$. The coordinate pair $(\theta,\phi)$ denotes both a point on the collection and on the trapping lens. A particle close to the focal point scatters the focused field.}
\label{fig:referenceSpheres}
\end{figure}

The situation under consideration is sketched in Fig.~\ref{fig:referenceSpheres}. An $x$ polarized plane wave is focused by a trapping lens with numerical aperture $\NA_{\rm tl} = \sin(\Theta_{\rm tl})$ before being recollimated by a collection lens with numerical aperture $\NA_{\rm cl} = \sin(\Theta_{\rm cl})$. A dipolar scatterer is positioned at $\vect{r}_0$ in close vicinity to the focus located at the origin. The scattered field, given by Eq.~\eqref{eq:dipole_scattered}, is interfered with the trapping beam on the reference surface of the collection lens.
We formulate the fields using the formalism laid out in Ref.~\onlinecite{Novotny2012}, such that the electric field on the reference sphere of the trapping lens reads
    \begin{equation}
    \label{eq:Eref_tr}
    \mathbf{E}_{\infty} = E_{\rm inc} \left[-\sin(\phi)\vect{n}_\phi + \cos(\phi) \vect{n}_\theta \right] \sqrt{\cos(\theta)}.
    \end{equation}
Here, $\vect{n}_\phi$ and $\vect{n}_\theta$ are spherical unit vectors along the azimuthal and polar direction, respectively.
Note that in our notation, the polar angle $\theta$ spans the range $[0\ldots \pi/2]$, while the azimuthal angle $\phi$ spans $[0\ldots 2\pi]$. 
Therefore, the coordinate pair $(\theta,\phi)$ denotes both a point on the trapping lens and a point on the collection lens (lying diametrically opposite relative to the origin).
In this notation, the trapping field on the collection lens (where it serves as the reference field in a forward-scattering detection scheme) takes the exact same form as the field in Eq.~\eqref{eq:Eref_tr}, besides an additional (irrelevant) phase factor accounting for propagation through the focal region.

\subsection{Quadrant detection in forward direction}\label{app:ForwardDetection}
We first turn our attention to the case of detection in forward direction, where the field trapping the particle naturally serves as a self-aligned reference field.
The total field on the reference sphere of the collection lens reads $\mathbf{E}_{\infty} + \imu \mathbf{E}_{\rm sc} $, with $\vect{E}_\mathrm{sc}$ the field scattered by the particle from Eq.~\eqref{eq:dipole_scattered}. The relative phase between both fields is fixed to $\pi/2$, since the field in the focal region (driving the dipolar scatterer) carries the Gouy phase shift relative to the field on the reference sphere (and the polarizability of the particle is assumed to be purely real)~\cite{Novotny2012}. 

Under the assumption of a strong reference field and a small particle displacement $\vect{r}_0$ from the origin, the differential power in direction $(\theta,\phi)$ per unit solid angle is
\begin{equation}\label{eq:diff_det_pwr_real}
    \begin{split}
    \frac{\diff p_\text{cl}(\theta,\phi)}{\diff \Omega} &= P_\text{inc} \cos(\theta) + \sqrt{3 P_\text{dip} P_\text{inc} \cos(\theta) /(2\pi)} \\
    &\times \left[\sin(\phi)^2+\cos(\phi)^2\cos(\theta)\right] \\
    &\times k\left(\mathbf{r}_0 \cdot \mathbf{n}_r - A z_0 \right). \\
    \end{split}
\end{equation}
Here, $P_\text{inc} \propto E_\text{inc}^2$ is the power of the trapping beam. 
Note that the polarization of the scattered dipole field $\vect{E}_{\rm dip} \propto [\vect{n}_x - (x/r) \vect{n}_r]$ differs from the reference field.
As usual in homodyne detection, we find two contributions to the power. The first contribution is independent of the scatterer's position and dominates the associated photon shot noise, while the second (interference) term is a measure for the scatterer's position $\vect{r}_0$.

Typically, the intensity distribution on the reference sphere is not spatially resolved. Instead, the fields are sent to a quadrant photodetector aligned along the $x$ and $y$ axes. The detector measures the integrated power striking the individual quadrants. The power $P^Q_1$ in the first quadrant is given by 
        \begin{equation}
        \label{eq:fw_quadrant_power}
        \begin{split}
        P^Q_1   =& \int_0^{\Theta_{\rm cl} } \diff\theta~\sin(\theta) \int_{0}^{\pi/2} \diff\phi ~p_\text{cl}(\theta,\phi)  \\
                = & P_{\rm inc} \frac{\pi}{4} \NA_{\rm cl}^2 + k\sqrt{3P_{\rm inc} P_{\rm dip} /(2\pi) } \vect{B}^\mathrm{fw} \mathbf{r}_0.
        \end{split}
        \end{equation}
The powers $P^Q_n$ in the other quadrants ($n=2,3,4)$ are equivalent with replacement $x_0 \to -x_0$ for $n\in\{2,3\}$ and $y_0\to-y_0$ for $n\in\{3,4\}$. In Eq.~\eqref{eq:fw_quadrant_power}, we have furthermore introduced the quantity
        \begin{equation}\label{eq:B_n_fw}
            \vect{B}^\mathrm{fw} = \int_0^{\Theta_{\rm cl} } \diff\theta ~ s\sqrt{c} \begin{pmatrix} s ( 1+2c)/3 \\ s ( 2+c)/3 \\ \pi (c - A)(1+c)/4 \end{pmatrix},
        \end{equation}
where we use the abbreviations $s=\sin(\theta)$ and $c=\cos(\theta)$. 
An analytical solution of the third component $B_z^\text{fw}$ is given in Appendix~\ref{app:analytical_solutions}.

We note that the particle position $\vect{r}_0$ can be extracted from the values $P_n^Q$ of the quadrant detector. Specifically
    \begin{subequations}\label{eq:sum_of_quadrantPowers}
    \begin{align}
       (P_1^Q + P_4^Q) - (P_2^Q + P_3^Q) = & P^{\mathrm{cal}}_x ~kx_0, \\
        (P_1^Q + P_2^Q) - (P_3^Q + P_4^Q)=& P^{\mathrm{cal}}_y ~ky_0, \\
        \sum _{n=1}^4 P^Q_n -P_\mathrm{inc}\pi\NA^2_\mathrm{cl}=& P^{\mathrm{cal}}_z ~kz_0,
    \end{align}
    \end{subequations}
with the calibration factors
    \begin{equation}
        P_j^{\mathrm{cal}} = \sqrt{24P_\mathrm{inc}P_\mathrm{dip}/\pi}~B_j^{\mathrm{fw}}, \quad j\in\{x,y,z\},
    \end{equation}
where $B_j^{\mathrm{fw}}$ is the $j$th vector component of Eq.~\eqref{eq:B_n_fw}.
To get access to $z_0$, a constant reference power has to be subtracted. 
Assuming that the detection is shot-noise limited, each of the position measurements will be subject to fluctuations with a white power spectrum
\begin{equation}
 S_{\rm PP}^Q = \frac{\hbar kc}{2} P_{\rm inc} \NA_{\rm cl}^2,
\end{equation}
arising from the first term in Eq.~\eqref{eq:fw_quadrant_power}. By dividing these fluctuations by the calibration factor $(P^{\mathrm{cal}}_j k)^2$, we find the imprecision noise spectral density for the motion of the particle along the $j$ axis
        \begin{equation}
        \label{eq:Simp_fw_real}
        \begin{split}
        S_\text{imp}^{j,\text{fw}} = &  \frac{\hbar c}{k P_{\rm dip}}  \frac{\pi \NA_{\rm cl}^2}{48 \left(B_j^{\mathrm{fw}}\right)^2}.
        \end{split}
        \end{equation}
As detailed in Sec.~\ref{sec:real_det_system}, we compare the calculated imprecision noise of the realistic detection system to the one obtained in the ideal case [Eqs.~\eqref{eq:Simp_all}] to obtain the measurement efficiencies in forward scattering which are plotted in Fig.~\ref{fig:referenceSpheres_det_eff}(b)
\begin{subequations}\label{eq:eta_group}
\begin{align}
\eta_x^{\rm fw} =&  \frac{30 \left(B_x^\mathrm{fw} \right)^2}{\pi^2 \NA_{\rm cl}^2}, \label{eq:eta_x_fw} \\
\eta_y^{\rm fw} =&  \frac{15 \left(B_y^\mathrm{fw}\right)^2}{\pi^2 \NA_{\rm cl}^2}, \label{eq:eta_y_fw}  \\
\eta_z^{\rm fw} =&  \frac{1}{1 +\frac{5}{2}A^2}  \frac{15 \left(B_z^\mathrm{fw}\right)^2}{\pi^2 \NA_{\rm cl}^2}.\label{eq:eta_z_fw}
\end{align}
\end{subequations}

Finally, we discuss the feature that $\eta_z^\mathrm{fw}$ plotted in Fig.~\ref{fig:referenceSpheres_det_eff}(b) vanishes for a symmetric setup. 
This feature can be understood by looking at $B_z^\mathrm{fw} = (\pi/4) \int_0^{\Theta_{\rm cl} } \diff\theta ~ s\sqrt{c}  (c - A)(1+c)$ from Eq.~\eqref{eq:B_n_fw}, with $c=\cos(\theta)$ and $s=\sin(\theta)$. The factor $[\cos(\theta) - A]$ in the integrand changes sign as the collection angle $\theta$ passes a certain critical value. 
For a symmetric setup ($\NA_\mathrm{tl}=\NA_\mathrm{cl}$), the integration over $\theta$ is truncated such that it strictly vanishes. This result can be obtained directly by plugging Eq.~\eqref{eq:A} into $B_z^\mathrm{fw}$ and assuming $\NA_\mathrm{tl}=\NA_\mathrm{cl}$.

Note that the derivation presented here tacitly assumed $\NA_\mathrm{cl}<\NA_\mathrm{tl}$, since when $\theta$ exceeds the maximum angle of the trapping beam $\Theta_\mathrm{tl} = \sin^{-1}(\NA_\mathrm{tl})$, the reference power drops to zero, such that there is no signal (but also no excess noise). For this reason, the detection efficiencies plotted in Fig.~\ref{fig:referenceSpheres_det_eff}(b) are constant for $\NA_\mathrm{cl}>\NA_\mathrm{tl}$.

\subsection{Quadrant detection in backward direction}\label{app:BackwardDetection}

To analyze the case of backscattering detection, we assume a reference field which has the same spatial distribution as the trapping field. The reference beam is then identical to Eq.~\eqref{eq:Eref_tr} by our choice of coordinates. 
Following the same derivations as in Sec.~\ref{app:ForwardDetection}, with the field scattered by the dipole expressed in the coordinate system given by the reference spheres, we find expressions for the imprecision noise spectral densities in backward scattering $S_\text{imp}^{j,\text{bw}}$ identical to Eq.~\eqref{eq:Simp_fw_real} under the substitutions $\Theta_\mathrm{cl}\rightarrow\Theta_\mathrm{tl}$, $\NA_\mathrm{cl}\rightarrow\NA_\mathrm{tl}$, as well as $\vect{B}^\mathrm{fw}\rightarrow\vect{B}^\mathrm{bw}$, with         \begin{equation}\label{eq:B_n_bw}
            \vect{B}^\mathrm{bw} = \int_0^{\Theta_{\rm tr} } \diff\theta ~ s\sqrt{c} \begin{pmatrix} s ( 1+2c)/3 \\ s ( 2+c)/3 \\ \pi (c + A)(1+c)/4 \end{pmatrix}.
        \end{equation}
Note that $\vect{B}^\mathrm{bw}$ differs from $\vect{B}^\mathrm{fw}$ only in the third vector component. Since both $A$ and $\vect{B}^\mathrm{bw}$ depend on $\Theta_{\rm tl}$, the $z$ component can be further simplified by inserting Eq.~\eqref{eq:A} for $A$, which yields
        \begin{equation}\label{eq:Bz_bw}
            B_z^\mathrm{bw} =\frac{\pi}{2} \int_0^{\Theta_{\rm tl}} \diff\theta~ s \sqrt{c} (1+c) c.
        \end{equation} 
An analytical solution of $B_z^\mathrm{bw}$ is given in Appendix~\ref{app:analytical_solutions}.
In analogy to forward scattering, we compare the obtained imprecision noise $S^{j,\mathrm{bw}}_\mathrm{imp}$ to the one for an ideal measurement given by Eqs.~\eqref{eq:Simp_all}, in order to compute the detection efficiencies in backscattering $\eta^\mathrm{bw}_j$, as plotted in Fig.~\ref{fig:referenceSpheres_det_eff}(c).\\

\section{Analytical solutions}\label{app:analytical_solutions}
In this appendix, we calculate analytical solutions of Eqs.~\eqref{eq:A}, \eqref{eq:B_n_fw}, and \eqref{eq:Bz_bw}. To ease our notation, we define
\begin{subequations}\label{eq:CD_integral}
\begin{align}
C(\Theta) &= \int_0^{\Theta}\diff\theta ~ s \sqrt{c} (1+c) , \\
D(\Theta) &= \int_0^{\Theta}\diff\theta ~ s \sqrt{c} (1+c) c.
\end{align}
\end{subequations}
We can solve both integrals analytically and find
\begin{subequations}\label{eq:CD}
\begin{align}
C(\Theta) &= 2\left(\frac{8}{15} - \frac{\cos(\Theta)^{3/2}}{3} - \frac{\cos(\Theta)^{5/2}}{5}\right) , \\
D(\Theta) &= 2\left( \frac{12}{35} - \frac{\cos(\Theta)^{5/2}}{5} - \frac{\cos(\Theta)^{7/2}}{7} \right).
\end{align}
\end{subequations}
This allows us to find solutions of the following integrals in terms of the functions $C(\Theta)$ and $D(\Theta)$
\begin{subequations}\label{eq:integrals_analytical}
\begin{align}
    A &= \frac{D(\Theta_{\rm tl})}{C(\Theta_{\rm tl})} , \\
    B_z^{\mathrm{fw}} &= \frac{\pi}{4} \left[D(\Theta_{\rm cl}) - A C(\Theta_{\rm cl})\right], \\
    B_z^{\mathrm{bw}} &= \frac{\pi}{2} D(\Theta_{\rm tl}).
\end{align}
\end{subequations}

\end{document}